\documentclass[12pt,a4paper,final]{iopart}

\usepackage{iopams}  
\usepackage{graphicx}
\usepackage{cite}
\usepackage[breaklinks=true,colorlinks=true,linkcolor=blue,urlcolor=blue,citecolor=blue]{hyperref}
\usepackage{epstopdf}
\newcommand{\doi}[2]{\href{http://dx.doi.org/#1}{#2}}

\begin{document}

\title[]{Comparative simulations of Fresnel holography methods for atomic waveguides}

\author{V A Henderson, P F Griffin, E Riis and A S Arnold$^{1}$}
\address{Dept.\ of Physics, SUPA, University of Strathclyde, Glasgow, G4 0NG, UK}
\ead{$^{1}$aidan.arnold@strath.ac.uk}

\begin{abstract}
We have simulated the optical properties of micro-fabricated Fresnel zone plates (FZPs) as an alternative to spatial light modulators (SLMs) for producing non-trivial light potentials to trap atoms within a lensless Fresnel arrangement. We show that binary (1-bit) FZPs with wavelength ($1\,\mu$m) spatial resolution consistently outperform kinoforms of spatial and phase resolution comparable to commercial SLMs in root mean square error comparisons, with FZP kinoforms demonstrating increasing improvement for complex target intensity distributions. Moreover, as sub-wavelength resolution microfabrication is possible, FZPs provide an exciting possibility for the creation of static cold-atom trapping potentials useful to atomtronics, interferometry, and the study of fundamental physics.

\end{abstract}

\section{Introduction}

Atom interferometry is a powerful tool for precise measurements and metrological technologies. It can be used for a wide range of applications, from the determination of fundamental constants and cosmological phenomena \cite{Burrage2015, Gupta2002} to navigation applications such as accelerometers and gyroscopes \cite{Barrett2014, Cronin2009, Gustavson1997}.  Developments in laser cooling, trapping and atom manipulation have allowed a wide range of atom interferometers to be developed \cite{Burrage2015, Gupta2002, Barrett2014, Cronin2009, Gustavson1997, Pritchard2012, Zawadzki2010, Arnold2006}, and for the exploration of light based atom traps \cite{Nshii2013, Schonbrun2008, McGilligan2015, Bruce2011, Trypogeorgos2013, Gaunt2012, Pasienski2008}.  Optical traps can offer a method of production for much more complex micrometer scale traps such as atomtronic optical circuits \cite{Seaman2007}.

Toroidal trapping of cold atoms for use as atom circuits has many applications beyond interferometry \cite{Halkyard2010, Marti2015}, such as the study of persistent currents in superfluids \cite{Ramanathan2011, Wright2013, Murray2013}, and low-dimensional atomic systems \cite{lowd,lowd2}. However, trapping ultra-cold atoms requires a very smooth trap, as the presence of very small perturbations in a potential can result in heating of a cold atom cloud or fragmentation of a trapped Bose-Einstein condensate \cite{Fortagh2002}. Within previous demonstrations of all-optical ring trapped BECs, the azimuthal variation of the ring minimum was far below the chemical potential of the BEC, with these rings produced through a variety of methods such as painted potentials \cite{Ryu2013} or combinations of confining light sheets with shaped light, for instance, Laguerre-Gaussian beams \cite{Ramanathan2011, Wright2013, Murray2013, Beattie2013}, co-axial focused beams \cite{Marti2015}, or conical refraction based beams \cite{Turpin2015}. To successfully produce trapping potentials for BEC, we must aim to match or surpass the above limit on azimuthal variation, thus aiming to produce traps of $\mu$K depth with a roughness of below 1\%.

There are many methods which can be used to produce tailored optical potentials, ranging from acousto-optic  beam deflection techniques \cite{Trypogeorgos2013, Henderson2009}  to holographic phase manipulation using a phase adjustable spatial light modulator (SLM) \cite{Bruce2011, Gaunt2012, Pasienski2008, Bruce2015} or digital micromirror device (DMD) \cite{Ha2015, Muldoon2012}.  To date, the holographic method has proved to be very adaptable, paving the way for the production of novel optical lattices for quantum simulation \cite{Nogrette2014}, dark spontaneous-force optical traps \cite{Radwell2013} and exotic Laguerre-Gauss modes \cite{Amico2005, Franke-Arnold2007, Arnold2012}. Despite these successes, SLM holography for atom trapping still remains an imperfect and computationally intensive technique, notwithstanding significant improvement in the iterative algorithms used \cite{Pasienski2008, Gaunt2012, Harte2014}. This is due to a combination of system aberrations, low spatial resolution, dead space between pixels, and the difficulty of creating an algorithm that converges on a solution suitable for atom trapping (i.e. smooth and without background light which could cause low loading rates or tunnelling out of the trap \cite{Gaunt2012}) without lowering light usage efficiency.

\begin{figure}[!b]
 \centering
 \begin{minipage}{.45\columnwidth}
 \includegraphics[width=\textwidth]{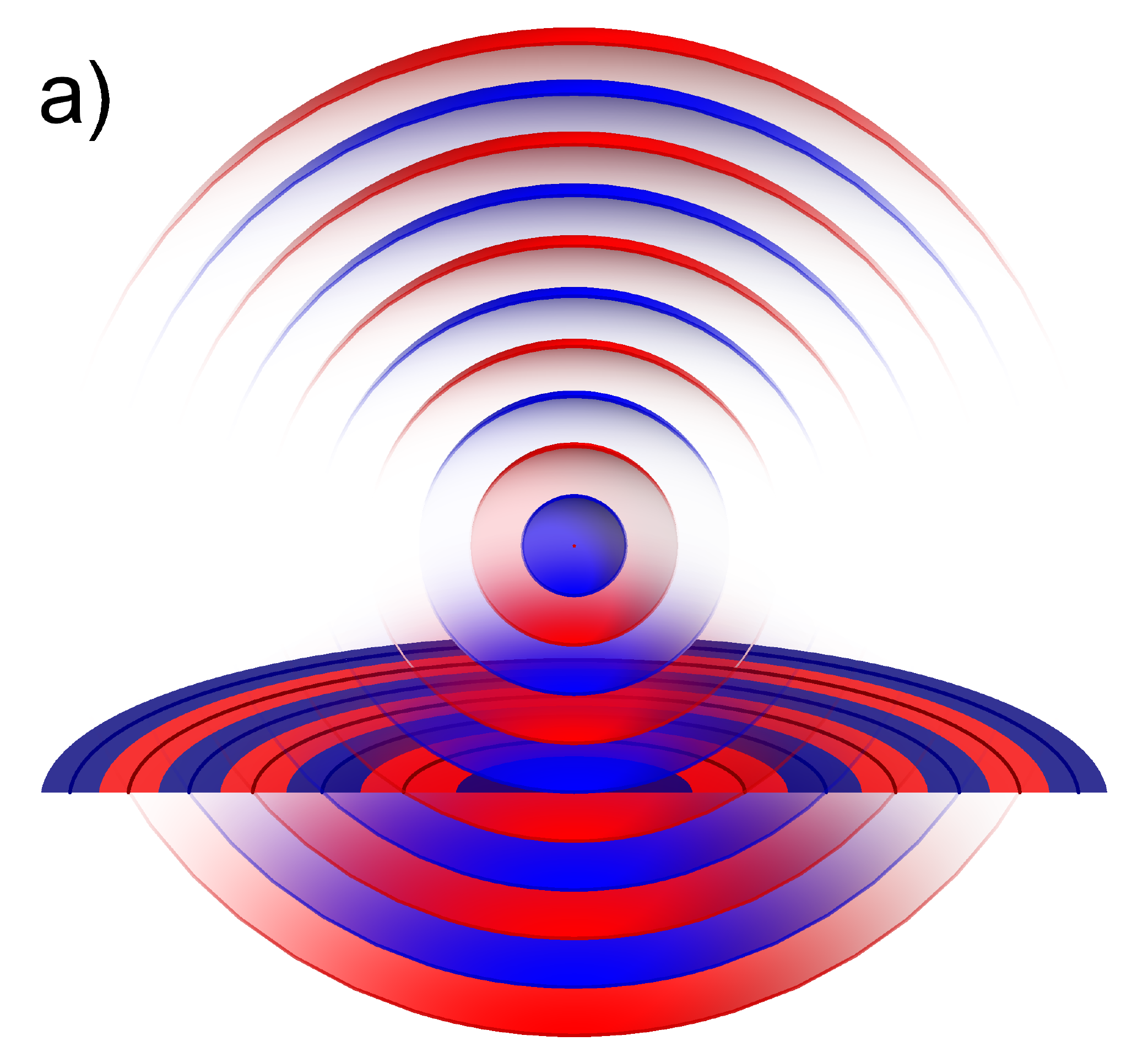}
\end{minipage}
\begin{minipage}{.45\columnwidth}
  \includegraphics[width= \textwidth]{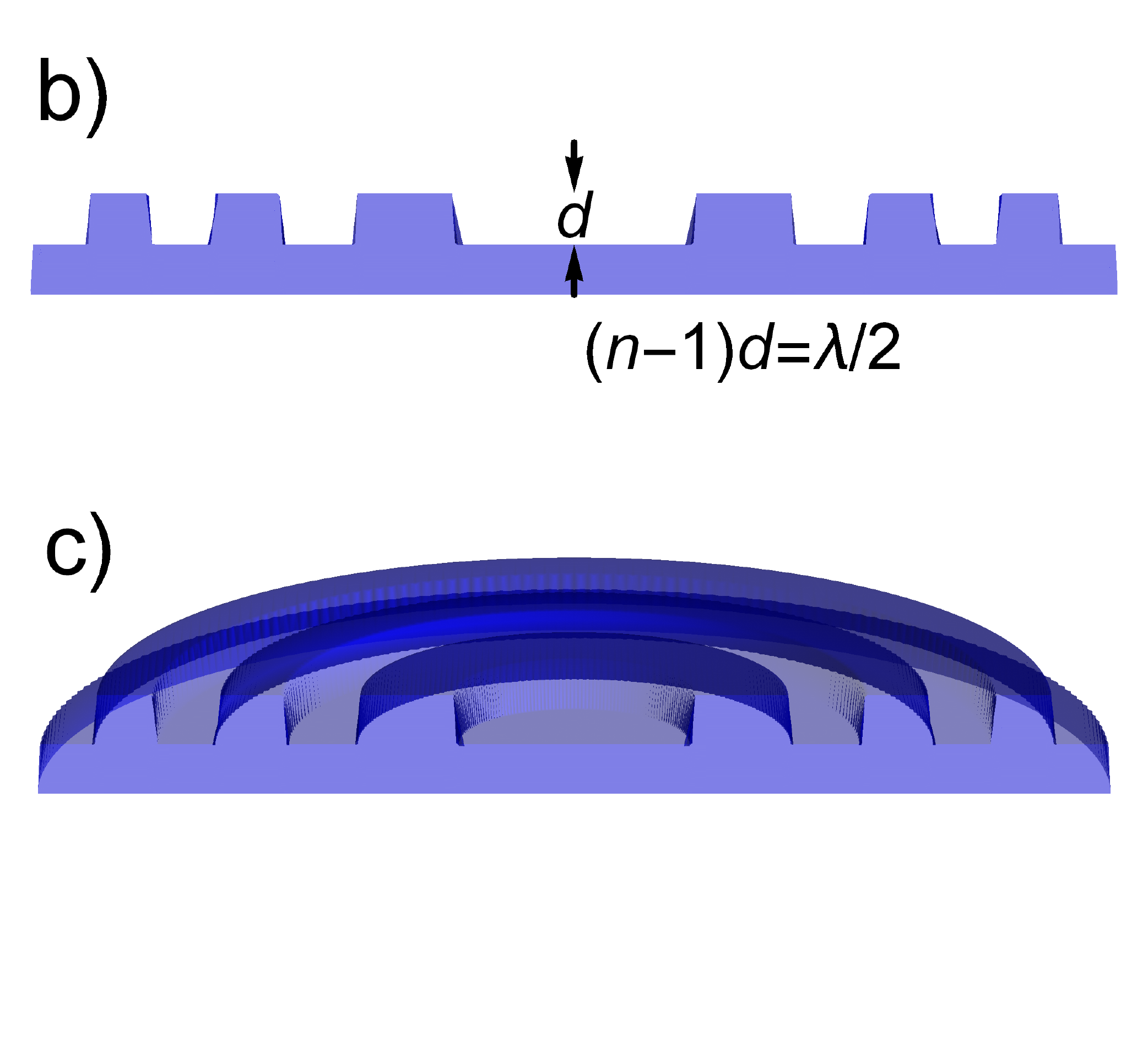}
  \end{minipage}
 \caption{a) Spherical light wave phasefronts (separated in phase by steps of $\pi$) emanating from a focused light beam form a distinctive Fresnel phase pattern when intersecting a plane.  b), c) Binary transmission holograms with equivalent phase characteristics are made from refractive index $n$ material, with half-wavelength steps in optical depth $(n-1)d$. Higher bit-depths of phase resolution enable hologram blazing.\label{fzp}}
\end{figure}

Fresnel Zone Plates (FZPs) work by spatially modulating either the amplitude or phase of a light beam, resulting in interference of the optical field after propagation; by design of the modulated region one can in principle then produce an arbitrary optical pattern, or trapping potential for atomtronics. The prototypical FZP is one that acts as a lens, resulting in a focused spot in the selected focal plane ($z = f$).  While the operation of such an FZP is standard in the teaching literature of diffraction, we find it intuitive to briefly consider the FZP required to generate a single focus, shown diagrammatically in figure~\ref{fzp} a). We make use of the time/direction symmetry of linear optics by starting from the desired result and finding the full electric field pattern at a defined plane. Our goal is now to create an optical element, the FZP, that matches an input beam, for example an idealised plane wave, to the field pattern that we produced in the plane. The FZP can then be considered the hologram generated by a plane wave and the backward-propagating field from the focus. For a binary FZP, we obtain a two--level map of the phases of the electric field in the plane of the FZP required to generate the desired focus. In the next section we discuss in detail the theory and numerical methods to implement this.  This type of plate (Fig.~\ref{fzp}) consists of alternating Fresnel zones forming concentric rings that alternate between the chosen binary states at radii,
\begin{equation}
r_j = \sqrt{j\lambda f + \frac{j^2 \lambda^2}{4}}~,
\label{eq:fres_r}
\end{equation}
where $j$ can take any integer value, and $\lambda$ is the wavelength of the incident light. Successive rings can be blocked, allowing only those that constructively interfere at the target plane to propagate. Alternatively, a phase shift of $\pi$ can be added to otherwise `destructive' zones, increasing the useful power at the focal plane. Figures \ref{fzp} b), c) demonstrate an envisaged transmissive binary FZP etched into a substrate, with consecutive zones that would be completely out of phase experiencing an increased optical path length.

\begin{figure}[!b]
 \centering
 \includegraphics[width=0.8\textwidth]{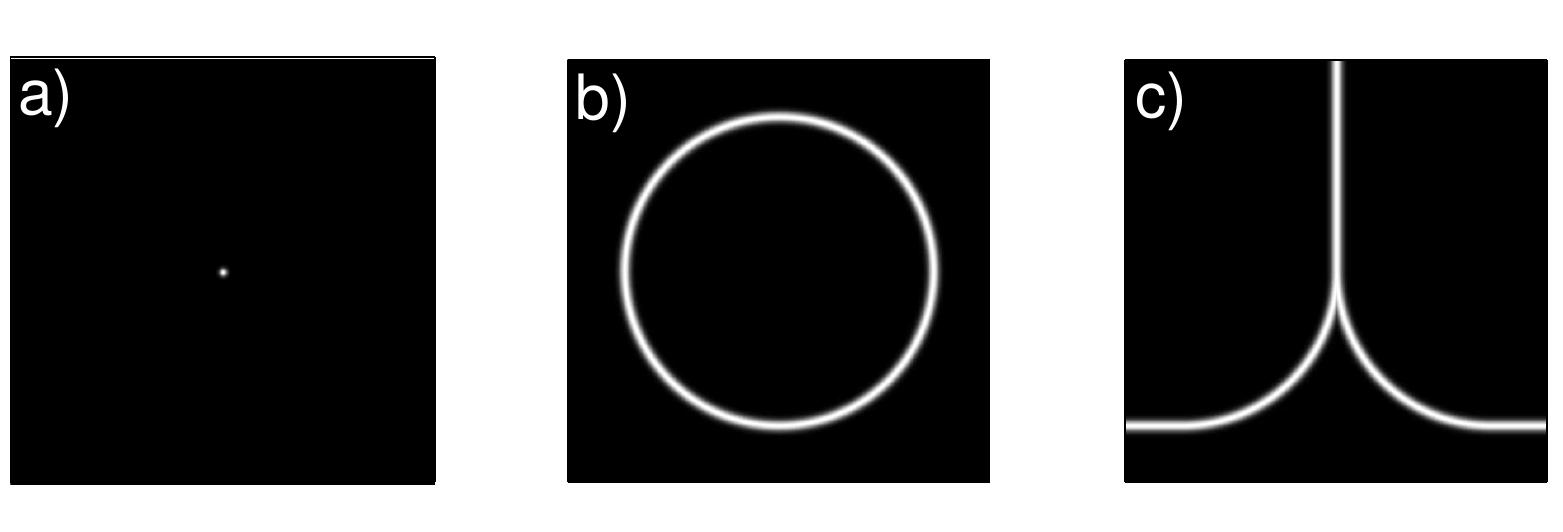}
 \caption{\label{fig:targets}The target intensity distributions used to simulate a range of potentials useful to atomtronics and interferometry; a), b), and c) show a focused spot, a ring and a beam-splitter, respectively. These simulation distributions are formed of Gaussians with $1/e^2$ widths of 2~$\mu$m (or 5~$\mu$m for the focus) and ring radii of $200~\mu$m, however, for visibility, the distributions shown above have a larger width and are cropped to show only the $600~\mu$m$\times 600~\mu$m area around the non-zero intensity.}
\end{figure}

A similar approach can be used to make straight waveguides with a linear symmetric FZP pattern, or to create arbitrary FZP-like patterns by recording the phase of a near-field diffraction pattern. In this work we will calculate and simulate phase plate patterns, or kinoforms, for single focii, rings, and beamsplitters, as shown in figure~\ref{fig:targets}. These target intensity distribution have been chosen due to their applicability to cold-atom trapping and atomtronics. The single focus allows both the calculation and propagation methods to be evaluated and compared to the simplest FZP model, whereas the ring allows for comparison of this method to existing toroidal traps which are the simplest nontrivial closed-loop circuits.  In order to extend the simulations to consider complex elements for atom optics we finally consider a beam splitter, as such an element is essential as a building block to create a circuit type interferometer.

We anticipate that microfabricated FZPs will overcome many of the limitations posed by the use of SLMs in atom trapping experiments. The higher spatial resolution and sharper edges between pixels presents the ability to reach higher spatial frequency and thus produce a wider range of more accurate holograms. Additionally, due to their size and transmissive operation, we expect that FZPs should be placed inside a vacuum chamber (as with the grating MOTs shown in \cite{Nshii2013, McGilligan2015}), thus immediately addressing the major system aberration of propagation through a vacuum chamber window. Further information will be discussed pertaining to the nature of FZPs in future sections.

\section{Simulation methods}
\begin{figure}[!b]
 \centering
 \includegraphics[width=0.5\textwidth]{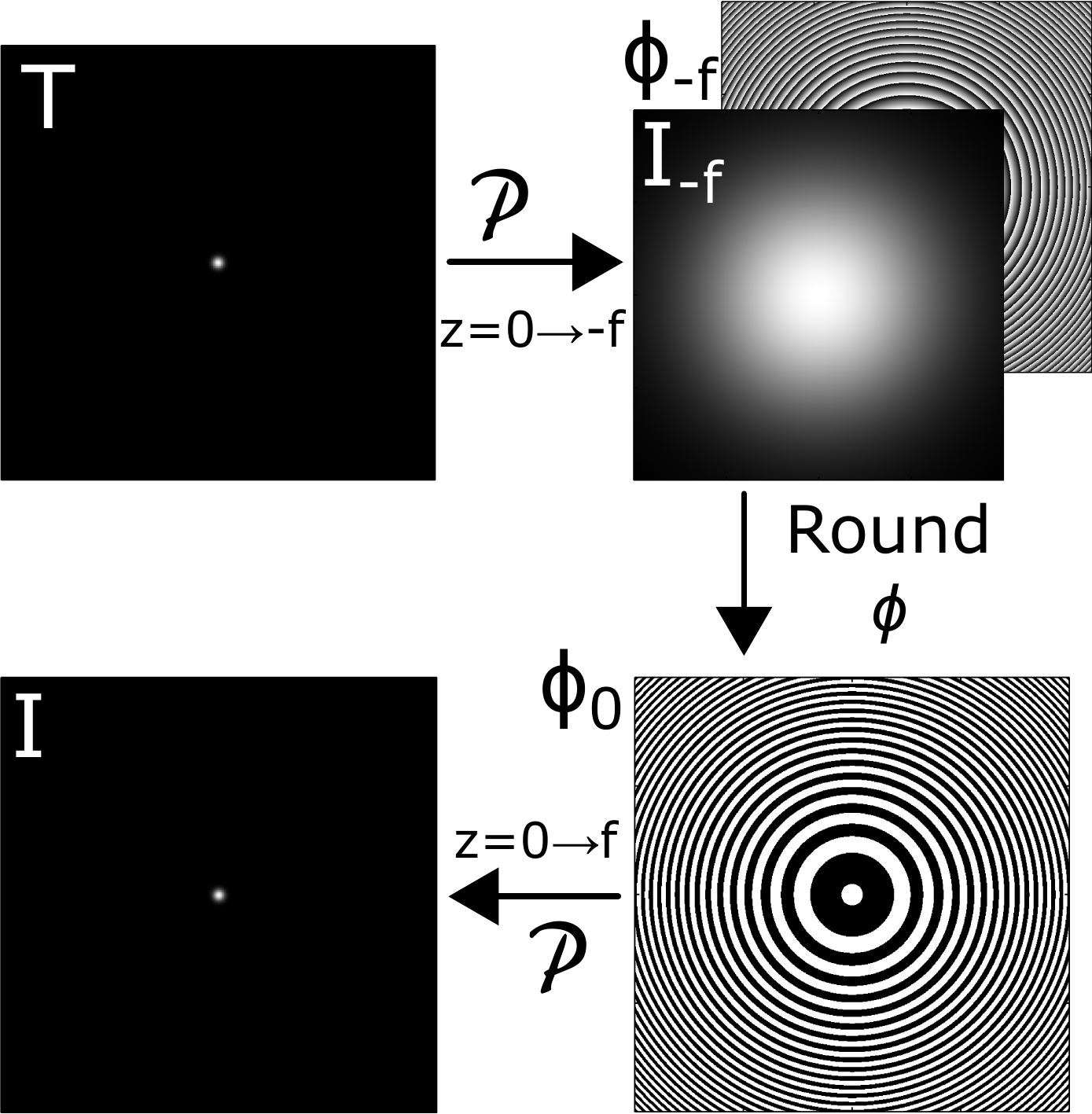}
 \caption{\label{fig:scheme} Schematic of the kinoform, or phase plate pattern, design process used. The target (T) electric field distribution is propagated backwards a distance $f$ using Fourier techniques and maximum spatial resolution (4096$\times$4096). The electric field is spatially averaged over a variable size (larger) grid of pixels, then separated into phase ($\phi_{\rm -f}$) and amplitude ($I_{\rm -f}$) components, with the phase rounded to 1-8 bit resolution. The kinoform is then illuminated to create an image.}
\end{figure}

The phase patterns required to produce the optical traps shown in figure \ref{fig:targets} are calculated using a Fourier--optics method of modelling the propagation of an initial electric field $\mathcal{E}^{(0)} = E(x', y', z=0)$ to a distance $z$. This uses the angular spectrum of the field, ($\mathcal{A}^{(0)}$), and the Helmholtz propagator, $\mathcal{H}$, such that,
\begin{equation}
\mathcal{E}^{(z)} = \mathcal{F}^{-1}\left [\mathcal{H}(z)  \mathcal{A}^{(0)} \right ] = \mathcal{F}^{-1}\left [e^{ik_z z}  \mathcal{F}\left [ \mathcal{E}^{(0)} \right ] \right ]~,
\label{eq:fourier_prop}
\end{equation}
where the z-component of the wave vector is $k_z = \sqrt{k^2 - k_x^2 - k_y^2}$ for an electric field with wave vector $k = 2 \pi / \lambda$ \cite{McDonald2015,Novotny2012}. We use this method, following the details in \cite{McDonald2015}, and references therein, to complete the design algorithm shown in figure \ref{fig:scheme}; firstly, a target intensity is calculated and then propagated backwards, using equation~\ref{eq:fourier_prop}, by the focal length. The phase of the resulting electric field in this plane is rounded to the desired bit depth, as discussed later in the text. This routine acts to calculate the required kinoform, and the performance of the result is tested numerically by simulating a desired input beam (either a plane wave or a Gaussian beam with defined width) that is then propagated forward by the focal length. Our method of simulation means that the pixel sizes of the kinoform and simulation (the electric field) are independent. Although we set the input beam and target plane to have flat phase fronts, we allow for phase freedom in the resultant distribution.  As we are not utilising a feedback algorithm, our method intrinsically avoids the presence of optical vortices, which can be confirmed through observations of simulation results. We consider the case in which the kinoform acts as a transmissive element and the incident light only illuminates the patterned area. It should also be noted that no optimisation is used to improve the kinoform. This full Helmholtz propagation method is computationally efficient and accurate, reducing the possibility of fringing artifacts in comparison to the paraxial approximation utilised in many hologram calculations as also highlighted in Ref.~\cite{Gaunt2012}.

To evaluate the success of each kinoform, we calculate the root mean squared (RMS) error for the normalised two dimensional final and target intensities,
\begin{equation}
\epsilon = \sqrt{\frac{1}{N} \sum \left ( \tilde{I}- \tilde{T}\right )^2}~,
\label{eq:rmserror}
\end{equation}
where $N$ is the number of pixels (in the simulation), $\tilde{I}$ is the final intensity, and  $\tilde{T}$ is the target intensity distribution, both intensity distributions are normalised by the mean of the pixels in $T$ that are brighter than $50\%$ of the maximum value \cite{Bruce2015}.

The target distributions we have chosen to simulate are shown in figure~\ref{fig:targets}: a) a focus with Gaussian waist (e$^{-2}$ radius) $w_0=5~\mu$m; b) a ring of radius $r=200~\mu$m and radial Gaussian waist $w_r=2~\mu$m; c) a beam splitter formed from straight segments and radii as given in b), again with waist  $w_b=2~\mu$m. 

Laser parameters of 2~mW (30~mW) power at a wavelength of 1064~nm were used for the focus (ring and beam splitter) simulations as these parameters give trap depths of a few $\mu$K. Moreover, trap frequencies are 2~kHz in the direction of tightest confinement, which is higher than existing ring shaped dipole potentials \cite{Marti2015, Wright2013, Murray2013, Eckel2014, Jendrzejewski2014, Ryu2013, Beattie2013} and permits access to lower dimensional regimes. The ring radius is larger than these previous demonstrations to increase its applicability to interferometry application where sensitivity scales with the area enclosed.  

Within the simulations, we run the calculations for a wide range of kinoform pixel sizes and phase resolution (or bit depth), allowing the comparison of binary FZP-type kinoforms with simulated pixel size of 1~$\mu$m to 8-bit SLM type kinoforms with simulated pixel sizes of 12~$\mu$m or more. The 12~$\mu$m  pixel size corresponds to the state of the art for SLMs, which have an effective area of approximately 2~cm$^2$, whereas FZPs can be manufactured with pixel size as small as 10~nm and with large total areas of up to 25~cm$^2$ \cite{Nshii2013}.  Despite these evident spatial advantages for FZPs, one must remember that SLMs typically operate with 8-bit precision and are updatable, whereas FZPs, by their very nature are static, with only two levels of phase control. Both technologies are already being utilised for trapping, in the form of optical tweezers \cite{Schonbrun2008,Thalhammer2011}.

Throughout the simulation process, the electric field propagation is calculated to a resolution of a wavelength with a simulation area of 4.38$\times$4.38~mm$^2$ ($2^{12}\,\lambda\times2^{12}\,\lambda$), limited solely by the reverse propagation technique and computation memory requirements.  For illumination by Gaussian beams, the choice of the input beam e$^{-2}$ radius, $w(z)$, is determined by the desired focal length and the Gaussian width, $w_0$ of the desired features by $w(z)=w_0\sqrt{1+(z/z_{\rm R})^2}$, with Rayleigh length $z_{\rm R}=\pi\,w_0^2/\lambda$. We do note that these computation limitations mean that the active area is smaller than, if comparable to, typical SLM active areas.

\section{Results \& Discussion}

\begin{figure}[!p]
 \centering
\includegraphics[width=.95\textwidth]{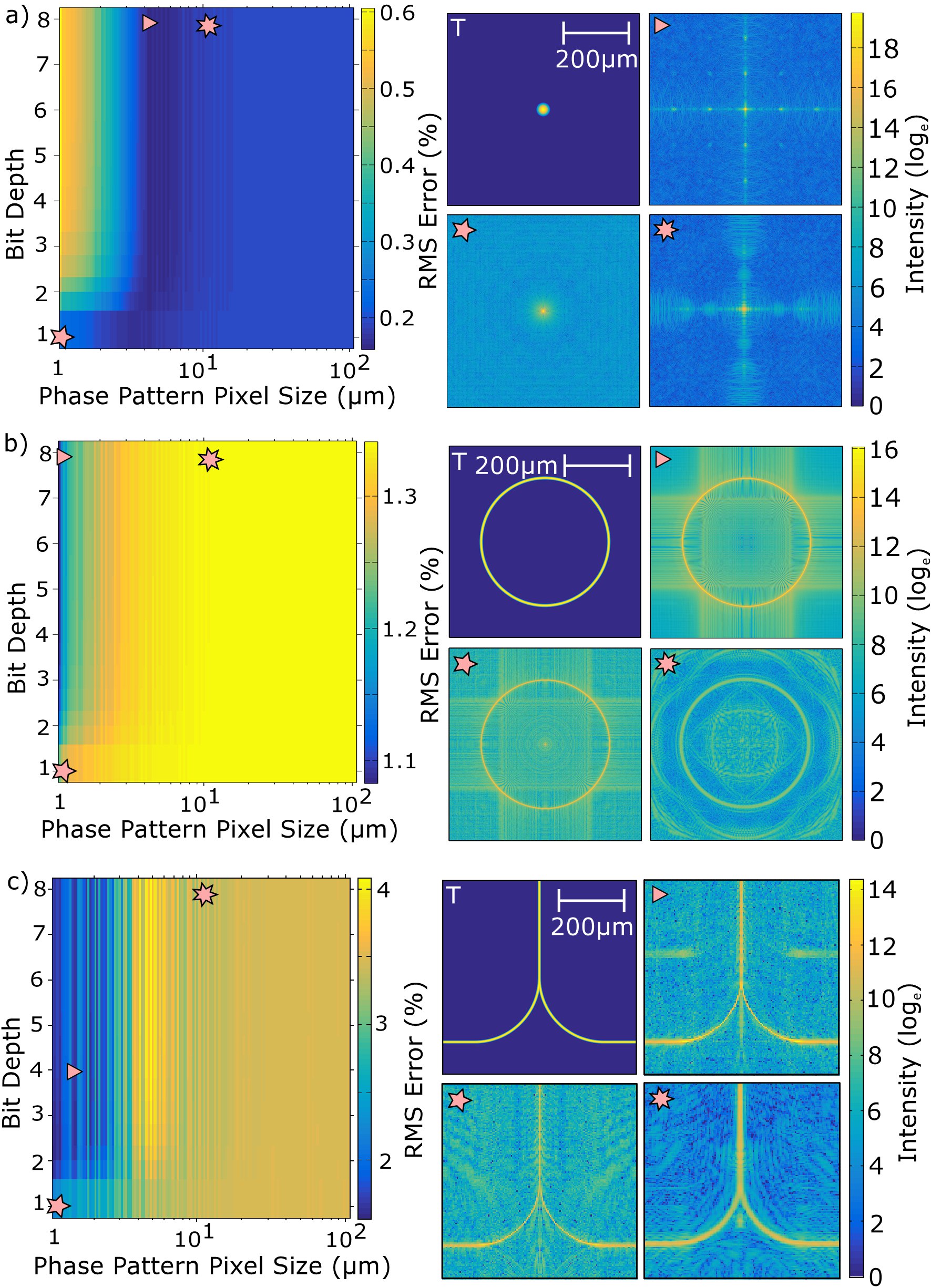}
 \caption{\label{fig:RMSErrorPlane} Plot of RMS error for kinoforms of varying spatial and phase resolution, illuminated by plane waves. The target intensity distributions, labelled T and with a scale bar, are to the right of the corresponding RMS error plot.   The obtained intensity distributions for the lowest RMS error, typical FZP, and typical SLM are labeled by the triangle, 5-point star and 7-point star respectively.}
\end{figure}

\begin{figure}[!p]
 \centering
\includegraphics[width=.85\textwidth]{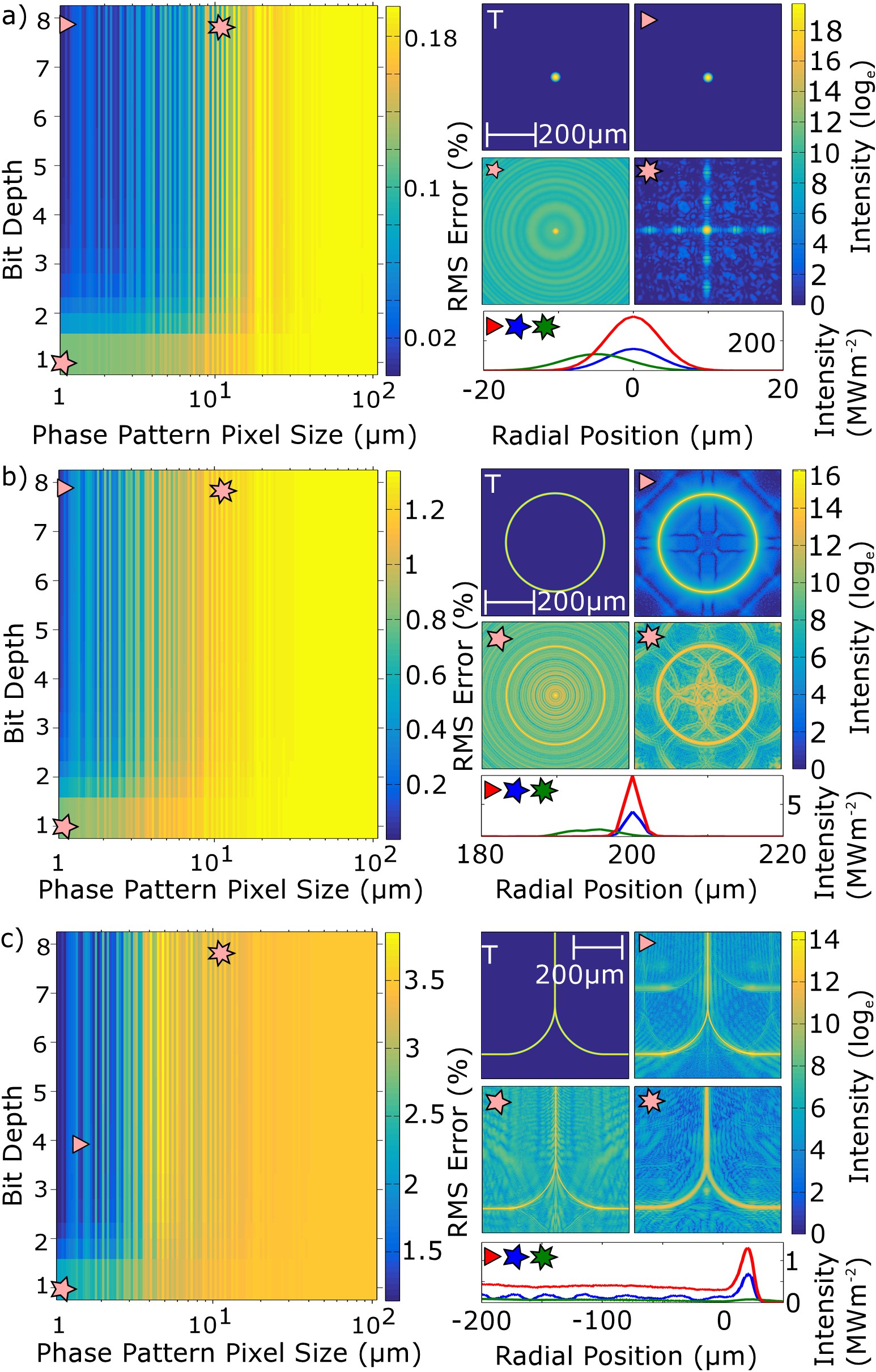}
 \caption{\label{fig:RMSErrorGaus} Plot of RMS error for kinoforms of varying spatial and phase resolution, illuminated by Gaussian beams of optimised widths. The obtained intensity distributions for the lowest RMS error, typical FZP, and typical SLM are labelled by the triangle, 5-point star and 7-point star respectively, shown logarithmically. Line graphs of intensity versus radial position is shown below the full intensity plots. For the focus and ring, the area around the (symmetrical) brightest region is shown at an appropriate scale. The equivalent for the beam splitter shows the intensity distribution along the vertical line of symmetry, with the peak offset from the distribution centre indicating the position of split.}
\end{figure}

\begin{figure}[!t]
 \centering
\includegraphics[width=.4\textwidth]{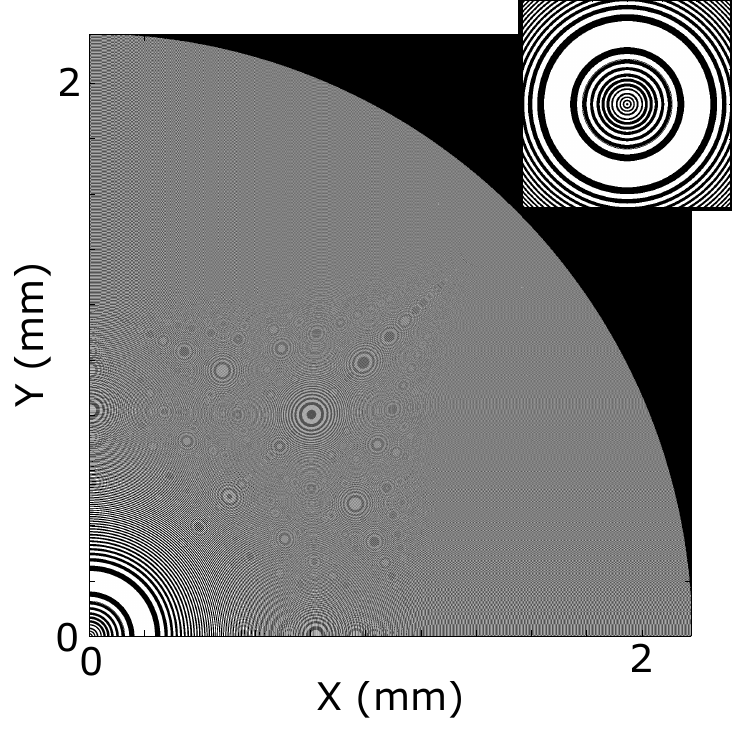}~~~~~~
 \includegraphics[width=.4\textwidth]{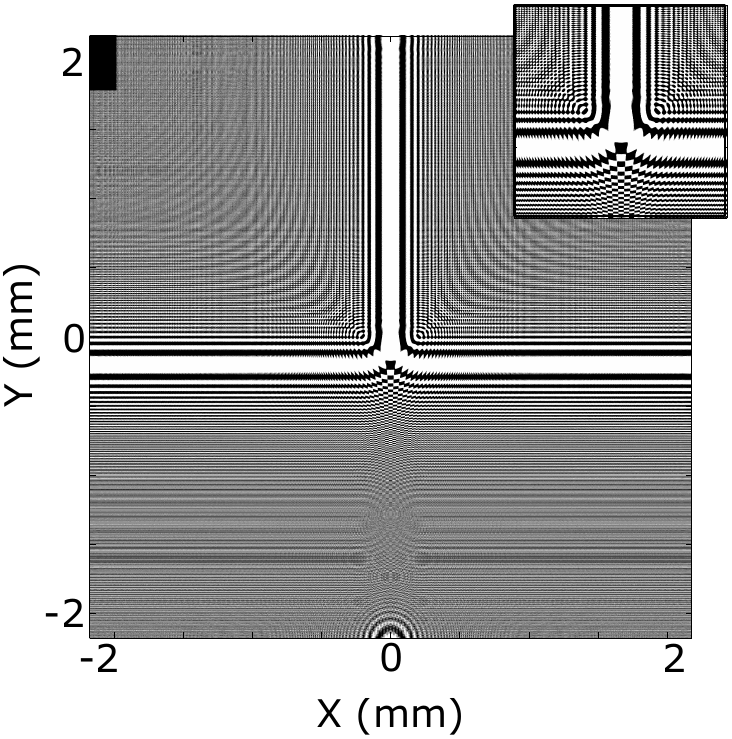}
 \caption{\label{fig:FZPexample} Fresnel Zone Plates calculated for producing a ring and a beam splitter using Gaussian beam illumination (as highlighted by the 5-point star in figure \ref{fig:RMSErrorGaus}).  The inset shows the central section of the kinoform, magnified to allow the zone plate features to be easily seen. Note that the outer regions of the zone plates appear grey due to pixel dithering where the Fresnel zones would be smaller than a pixel. The pure black area denotes the masked area, where the plate is non-transmissive or light is blocked. The off-centre appearance of rings in the ring kinoform are artefacts of the finite simulation pixel size.}
\end{figure}

Maps of RMS error, calculated in the simulations using equation~\ref{eq:rmserror}, are shown in figures \ref{fig:RMSErrorPlane} and \ref{fig:RMSErrorGaus}.  For all three target patterns and illumination beams (except the plane focus), there is a clear increase in RMS error with increasing pixel size and decreasing bit depth.  The simulations also show that a two level FZP consistently has an RMS error lower than that of a kinoform comparable to an SLM. In addition, we can note that, at low pixel size, increasing the bit depth from 2 to 4 level phase resolution significantly reduces the RMS error, thus improvements in microfabrication techniques would significantly increase the accuracy of the FZP kinoforms by allowing for non-binary phase.  Examples of the calculated kinoforms for FZPs illuminated by a Gaussian and producing a ring and beam splitter are shown in figure \ref{fig:FZPexample}.

The RMS error map, shown in figure~\ref{fig:RMSErrorPlane}, for a focus kinoform illuminated by a plane wave clearly shows an unexpected increase in RMS error at high phase (high bit depth) and spatial resolution (small pixel size).  In this area of higher RMS error, we observe that the optical power is concentrated in a tighter focus than the target 5~$\mu$m e$^{-2}$ radius.  This can be understood because each pixel of the kinoform is illuminated equally, unlike Gaussian optics where a concomitant Gaussian illumination of the optical element is required.  We can explore the consequences of this by considering three of the contributors to the RMS error: phase resolution error, spatial resolution error, and illumination error.  In our algorithm all spatial intensity information from back-propagation is lost and replaced by the intensity information of the illumination beam, whereas the phase information loss is only limited by the pixel size and phase resolution.  At large pixel size and low phase resolution, these sources of error dominate over the intensity error, but at high resolutions, the lost intensity information becomes dominant.  It is a standard result in Gaussian optics that a smaller focus diverges more rapidly than a larger focus, meaning that the tighter the focus desired, the larger a kinoform or lens should be used, such that the numerical aperture can be increased. Conversely, this means that the size of the illuminated area of the kinoform, rather than the phase across it, affects the size of the focus produced. So, for the plane wave case, the illumination is more similar to that required for a smaller focus than 5~$\mu$m.  We do not see this in the Gaussian illumination simulations, figure~\ref{fig:RMSErrorGaus}, due to the Gaussian weighting of the intensity at the kinoform.

As one can see from the RMS errors shown in figures \ref{fig:RMSErrorPlane} and \ref{fig:RMSErrorGaus}, accuracy of intensity reproduction is reduced with pattern complexity and for distributions with less obvious symmetries: reproduction of the beam splitter is much less accurate than for either the focus and the ring.  Both the ring and the focus have been masked to form circularly symmetric kinoforms, meaning that artefacts caused by the square shape of the active area are reduced, however, the reduced symmetry of the beam splitter makes this process more complex. The masking makes pixels outside of a desired area completely dark, thus creating an ‘active area’ of illuminated pixels and excluding pixels which cause abberations. In the beam-splitter case, we were able to use the symmetry properties of a straight waveguide Fresnel zone plate to shape the active area appropriately, thus blocking light incident more than a certain distance from centre of the intensity lines. This technique greatly improved accuracy, but requires further fine-tuning to allow the approach to be applied to an arbitrary intensity pattern.  However, we note that if the appropriate spatial distribution of the incident field, with a flat phase front, can be produced at the kinoform, then the errors would rapidly tend to zero, as for the single focus in the upper plots of figure~\ref{fig:RMSErrorGaus}. Indeed, producing such a large scale pattern is well suited to the coarser resolution of an SLM, suggesting that SLMs and FZPs can be used together synergistically. 

In all the error maps, particularly for the Gaussian illumination, we see non-monotonic variations in the errors between consecutive pixel sizes. This is due to aliasing between the three length scales involved in the kinoform design calculations: the length scale of phase change, the simulation pixel size ($\lambda$), and the kinoform pixel size. Due to the involvement of three length scales we were not able to reduce this roughness with suitable choice of any of these values. The roughness in RMS error is less pronounced for plane wave illumination as the overall RMS error is higher and so this aliasing is less prominent.

We can also note that the discontinuous nature of the example beam splitter has also increased the error in its production, this led to us using a target that reached the edges of the simulation area to avoid such issues. In a useful intensity distribution for atomtronics, one would want to produce a target intensity with no discontinuities (i.e. a closed-loop circuit), such as a ring with a beam splitter at either end for use in interferometry; hence, the discontinuity based artifacts and errors are not critical to the success of these simulations.  

\begin{figure}[!b]
 \centering
 \includegraphics[width=0.48\textwidth]{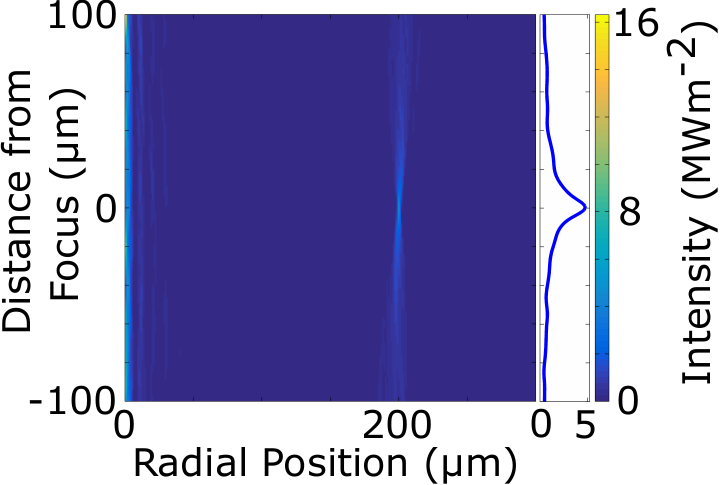}~~~~~~\includegraphics[width=0.48\textwidth]{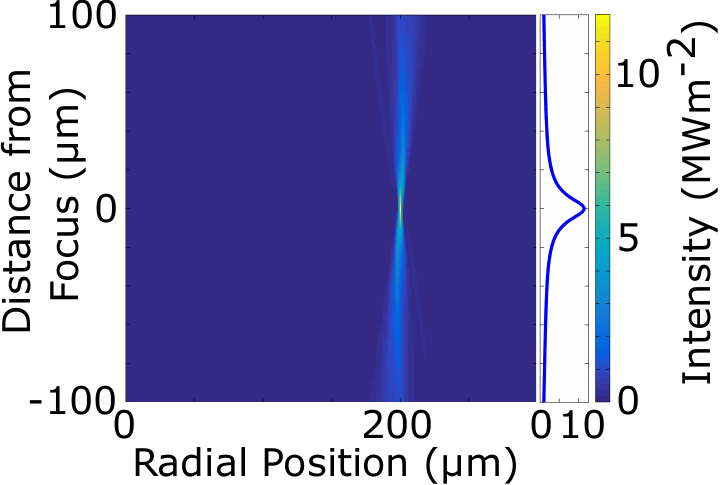}
 \caption{\label{fig:Gausringprop} Propagation through the focus for a ring hologram generated using an FZP (left) and the best kinoform (right). The average intensity of points within 0.5$\mu$m of the ring radius is shown in the cross-section plots  on the right of each image.}
\end{figure}

In order to demonstrate the applications of the hologram method of optical trap generation (particularly the potential for three dimensional trapping), we have demonstrated propagation through the focus of the ring distribution in figure \ref{fig:Gausringprop}. This is shown both for the best kinoform and for an FZP, with the average intensity of the ring minimum at each distance shown as a scatter plot alongside the full intensity distributions. Both cases demonstrate a full-width-half-maximum (FWHM) in the propagation direction of 20~$\mu$m, similar to that expected for a focussed Gaussian beam. Azimuthal plots of intensity are shown as line graphs in figure \ref{fig:RMSErrorGaus}, allowing for intensity noise to be seen. We can note that the intensity distribution in the case of the focus and ring are too narrow to show any noise due to the pixel size of the simulation, however, we can see significant noise along the vertical waveguide section of the beam splitter. The beam splitter is noise is largely due to beating between the vertical and horizontal sections of the waveguides and could be minimised with more careful target distribution design.

\begin{figure}[!t]
 \centering
 \includegraphics[width=0.48\textwidth]{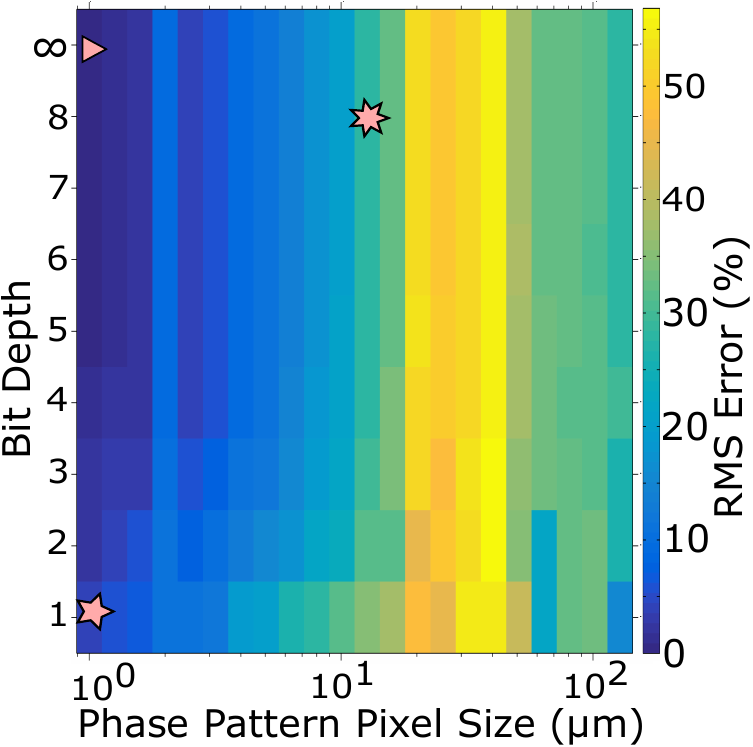}~~~ \includegraphics[width=0.48\textwidth]{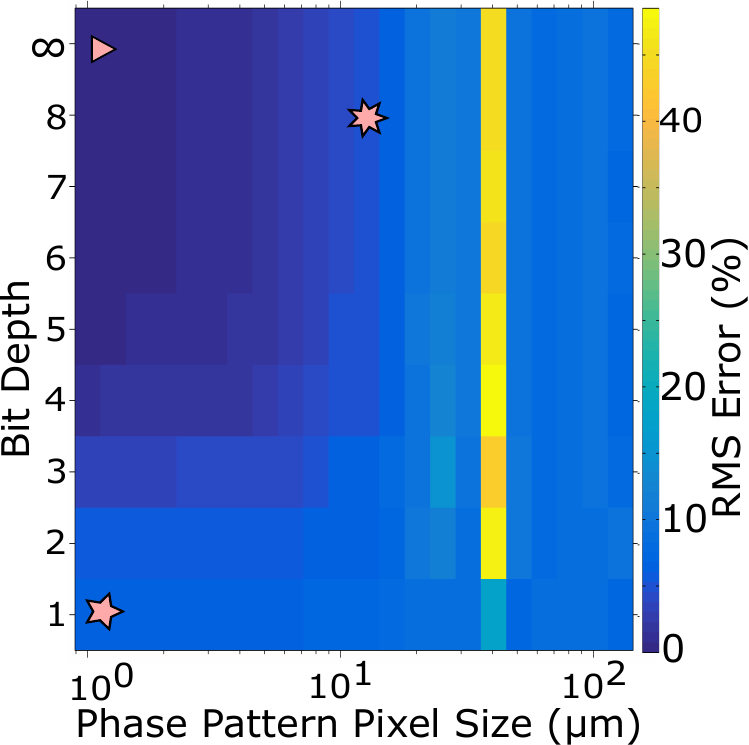}
 \caption{\label{fig:targetweighted} For the ring-shaped target, image (a) is a demonstration of how the RMS error is modified if one considers only the grid points in which the target is within $50\%$ of the maximum intensity. The target is normalised to its maximum value within this pixel range, and the image is scaled by a constant which  minimises the RMS error. Note the much higher overall error, as the large background content of the image can give a false impression of pattern smoothness.  The lowest RMS error, typical FZP, and typical SLM are labeled by a triangle, 5-point star and 7-point star, with corresponding errors of $0.0\%,$ $3.7\,\%$ and $32.7\%$, respectively. In image (b), for benchmarking, we consider a complex target `OR' gate ($150\times 280\, \mu$m$^2$) similar to that used in Refs.~\cite{Gaunt2012,Pasienski2008}. Note that in this case the lowest RMS error, typical FZP, and typical SLM are labeled by a triangle, 5-point star and 7-point star, with corresponding errors of $0.0\%,$ $17.4\,\%$ and $19.0\%$, respectively. Such values appear high, however it is important to consider the small target size, and that there is no additional hologram optimisation. The phase profile across the target is flat in all cases, with no observable vortices.}
\end{figure}

In the simulations of RMS errors in Figures~\ref{fig:RMSErrorPlane},  \ref{fig:RMSErrorGaus} we adopted a compromise position whereby we compared both target and image distributions across the whole grid size. This means that even the background wings (i.e.\ non-target zone) of the intensity distribution - which could affect the atomtronic circuit loading efficiency - contribute to the error. However, for a given application one may be mainly interested in a subset of the image and target, e.g. the pixel region where the top 50\% of the target intensity distribution. This region is where the coldest atoms would be trapped and in this case it makes sense to modify equation \ref{eq:rmserror} to only consider pixels in this zone. Moreover, one should then adapt  $\tilde{I}$ the final intensity, and $\tilde{T}$ the target intensity distribution, so that the intensity distributions are independently normalised by their maximum value over the pixels in $T$ which are brighter than $50\%$ of the maximum value. This gives a more realistic estimate of the in-situ trap roughness, which can be seen in figure \ref{fig:targetweighted}. The lowest RMS error, typical FZP, and typical SLM have corresponding errors of $0.0\%,$ $3.7\,\%$ and $32.7\%,$ respectively for a ring shaped target. In this situation, rather than the plane wave/Gaussian illumination considered in Figs.~\ref{fig:RMSErrorPlane} and \ref{fig:RMSErrorGaus}, the hologram is illuminated by its ideal spatial intensity distribution -- a realistic assumption we elucidate on in our conclusions.

\section{Outlook and Conclusion}

By calculating and simulating kinoforms for focii, rings and beam splitters, we have shown that under most circumstances spatial resolution is much more critical than the bit-depth of the hologram. Specifically, we demonstrated that, in the lensless Fresnel regime, FZPs with wavelength spatial resolution consistently show improved root mean square error over kinoforms of spatial and phase resolution comparable to commercial SLMs which are typically 8—bit, with 12~$\mu$m pixels. FZP kinoforms become increasingly superior for complex target intensity distributions, indicating their suitability for use to produce static atomtronic circuits for trapping ultracold atoms. This is accompanied by the illustration of 3D trapping capabilities through propagation of a ring shaped potential through its focus.  By extension of a FZP from a binary kinoform to a 4 level kinoform, the fidelity of intensity distributions can be greatly increased, showing the potential of these kinoforms to improve with increasing micro-fabrication capabilities \cite{Harvey2014}.

Despite the success of these simulations, they are limited to wavelength resolution due to the Helmholtz propagation method used and to a size of 4.38$\times$4.38~mm$^2$ by the memory requirements of the simulation.  The calculation process explicitly does not include an algorithm for iterative error correction of the kinoform meaning that both FZP and SLM RMS errors may find improvements with the use of algorithms similar to those used in \cite{Bruce2015, Gaunt2012}.

Future work will extend this method of kinoform calculation to include an optimisation algorithm, whilst the manufacture of potentially useful FZPs will allow for predictions to be tested experimentally. There is also great potential for combining the strengths of different techniques: a laser incident on a DMD pattern could be re-imaged onto the FZP in order to provide flat-phase front spatially-tailored intensity illumination (assumed in Fig.~\ref{fig:targetweighted}), with RMS errors substantially reduced beyond those from plane wave/Gaussian illumination. Whilst only red-detuned (bright) dipole potentials were considered in this paper, extension to blue-detuned (dark) traps should be straightforward, however such patterns rely more heavily on destructive interference which is likely to impinge on the smoothness of the final patterns. Additionally, FZPs should lend themselves to future extension work involving multi-wavelength hologram production following a similar approach to that shown in \cite{Bowman2015}.

For the dataset associated with this paper please see \cite{doi}.

\section*{Acknowledgments}

This work has been supported by DSTL, the Leverhulme Trust RPG-2013-074 and by the EPSRC as part of the Quantum Technology Hub for Sensors and Metrology. We are grateful to Jonathan Pritchard for initial assistance with the simulations.

\end{document}